\documentclass[aps,prc,twocolumn,superscriptaddress]{revtex4-1}

\usepackage{graphicx}
\usepackage{epstopdf}
\usepackage{multirow}
\usepackage[pagebackref=false]{hyperref}	% clickable pdf links
\hypersetup{
  colorlinks = true,
  linkcolor=blue,   % color of internal links
  citecolor=blue,   % color of links to bibliography
  urlcolor=blue,    % color of external links
  pdfdisplaydoctitle=true
}

\begin{document}

%\preprint{}

\title{Shape coexistence in neutron-deficient Hg isotopes studied via lifetime measurements in $^{184,186}$Hg and two-state mixing calculations}


% 1st author
\author{L.~P.~Gaffney}
\email[Corresponding author ($^{184}$Hg): ]{Liam.Gaffney@fys.kuleuven.be}
%\email[Corresponding author: ]{Liam.Gaffney@fys.kuleuven.be}
\affiliation{Oliver Lodge Laboratory, University of Liverpool, Liverpool L69 7ZE, United Kingdom}
\affiliation{KU Leuven, Instituut voor Kern- en Stralingsfysica, B-3001 Leuven, Belgium}
% 2nd author
\author{M.~Hackstein}
\email[Corresponding author ($^{186}$Hg): ]{hackstein@ikp.uni-koeln.de}
\affiliation{Institut f\"ur Kernphysik, Universit\"at zu K\"oln, Z\"ulpicher Str. 77, D-50937 K\"oln, Germany}
% 3rd author
\author{R.~D.~Page}
\email[Corresponding author (calc.): ]{rdp@ns.ph.liv.ac.uk}
\affiliation{Oliver Lodge Laboratory, University of Liverpool, Liverpool L69 7ZE, United Kingdom}
% 4th author
\author{T.~Grahn}
\affiliation{Oliver Lodge Laboratory, University of Liverpool, Liverpool L69 7ZE, United Kingdom}
\affiliation{Department of Physics, University of Jyv\"askyl\"a, P.O. Box 35, FI-40014 Jyv\"askyl\"a, Finland}
% 5th author
\author{M.~Scheck}
\affiliation{Oliver Lodge Laboratory, University of Liverpool, Liverpool L69 7ZE, United Kingdom}
\affiliation{Institut f\"ur Kernphysik, TU Darmstadt, D-64289 Darmstadt, Germany}
% 6th author
\author{P.~A.~Butler}
\affiliation{Oliver Lodge Laboratory, University of Liverpool, Liverpool L69 7ZE, United Kingdom}
% other authors, alphabetical
\author{P.~F.~Bertone}
\affiliation{Physics Division, Argonne National Laboratory, Argonne, Illinois 60439, USA}
\author{N.~Bree}
\affiliation{KU Leuven, Instituut voor Kern- en Stralingsfysica, B-3001 Leuven, Belgium}
\author{R.~J.~Carroll}
\affiliation{Oliver Lodge Laboratory, University of Liverpool, Liverpool L69 7ZE, United Kingdom}
\author{M.~P.~Carpenter}
\affiliation{Physics Division, Argonne National Laboratory, Argonne, Illinois 60439, USA}
\author{C.~J.~Chiara}
\affiliation{Physics Division, Argonne National Laboratory, Argonne, Illinois 60439, USA}
\affiliation{Department of Chemistry and Biochemistry, University of Maryland, College Park, Maryland 20742, USA}
\author{A.~Dewald}
\affiliation{Institut f\"ur Kernphysik, Universit\"at zu K\"oln, Z\"ulpicher Str. 77, D-50937 K\"oln, Germany}
\author{F.~Filmer}
\affiliation{Oliver Lodge Laboratory, University of Liverpool, Liverpool L69 7ZE, United Kingdom}
\author{C.~Fransen}
\affiliation{Institut f\"ur Kernphysik, Universit\"at zu K\"oln, Z\"ulpicher Str. 77, D-50937 K\"oln, Germany}
\author{M.~Huyse}
\affiliation{KU Leuven, Instituut voor Kern- en Stralingsfysica, B-3001 Leuven, Belgium}
\author{R.~V.~F.~Janssens}
\affiliation{Physics Division, Argonne National Laboratory, Argonne, Illinois 60439, USA}
\author{D.~T.~Joss}
\affiliation{Oliver Lodge Laboratory, University of Liverpool, Liverpool L69 7ZE, United Kingdom}
\author{R.~Julin}
\affiliation{Department of Physics, University of Jyv\"askyl\"a, P.O. Box 35, FI-40014 Jyv\"askyl\"a, Finland}
\author{F.~G.~Kondev}
\affiliation{Nuclear Engineering Division, Argonne National Laboratory, Argonne, Illinois 60439, USA}
\author{P.~Nieminen}
\affiliation{Department of Physics, University of Jyv\"askyl\"a, P.O. Box 35, FI-40014 Jyv\"askyl\"a, Finland}
\author{J.~Pakarinen}
\affiliation{Oliver Lodge Laboratory, University of Liverpool, Liverpool L69 7ZE, United Kingdom}
\affiliation{Department of Physics, University of Jyv\"askyl\"a, P.O. Box 35, FI-40014 Jyv\"askyl\"a, Finland}
\affiliation{CERN-ISOLDE, CERN, CH-1211 Geneva 23, Switzerland}
\author{S.~V.~Rigby}
\affiliation{Oliver Lodge Laboratory, University of Liverpool, Liverpool L69 7ZE, United Kingdom}
\author{W.~Rother}
\affiliation{Institut f\"ur Kernphysik, Universit\"at zu K\"oln, Z\"ulpicher Str. 77, D-50937 K\"oln, Germany}
\author{P.~Van Duppen}
\affiliation{KU Leuven, Instituut voor Kern- en Stralingsfysica, B-3001 Leuven, Belgium}
\author{H.~V.~Watkins}
\affiliation{Oliver Lodge Laboratory, University of Liverpool, Liverpool L69 7ZE, United Kingdom}
\author{K.~Wrzosek-Lipska}
\affiliation{KU Leuven, Instituut voor Kern- en Stralingsfysica, B-3001 Leuven, Belgium}
\author{S.~Zhu}
\affiliation{Physics Division, Argonne National Laboratory, Argonne, Illinois 60439, USA}


\date{\today}

% Abstract
\begin{abstract}
The neutron-deficient mercury isotopes, $^{184,186}$Hg, were studied with the Recoil Distance Doppler-Shift (RDDS) method using the Gammasphere array and the K\"{o}ln Plunger device. The Differential Decay Curve Method (DDCM) was employed to determine the lifetimes of the yrast states in $^{184,186}$Hg. An improvement on previously measured values of yrast states up to $8^{+}$ is presented as well as first values for the $9_{3}$ state in $^{184}$Hg and $10^{+}$ state in $^{186}$Hg. $B(E2)$ values are calculated and compared to a two-state mixing model which utilizes the variable moment of inertia (VMI) model, allowing for extraction of spin-dependent mixing strengths and amplitudes.
\end{abstract}


%\keywords{}

\maketitle

\section{Introduction}
Nuclei exhibiting different shapes at low energy have been of interest in nuclear structure ever since the discovery of a large jump in the mean-squared charge radius, associated with a dramatic change in shape between $^{187}$Hg and $^{185}$Hg observed in isotope shift measurements~\cite{Bonn1972}. Calculations based on Strutinsky's shell-correction method~\cite{Frauendorf1975} interpreted this result as a transition from a weakly-deformed oblate to a more pronounced prolate-deformed shape. Further isotope shift measurements reveal that the weakly-deformed oblate character extends down to $A=182$ in the even-mass Hg isotopes~\cite{Ulm1986}. Calculations in these isotopes using the Nilsson-Strutinsky approach~\cite{Nazarewicz1993} predict two deformed minima, where the lowest-energy minimum corresponds to an oblate shape, $\beta\simeq-0.15$, and the second to a more deformed prolate shape with $\beta\simeq0.27$. In shell-model terms, these minima are associated with a proton zero-particle-two-hole configuration, $\pi(0p-2h)$, and a two-proton excitation across the $Z=82$ shell gap yielding a $\pi(2p-4h)$ configuration, respectively.

\begin{figure}[tb]
\includegraphics[width=\columnwidth]{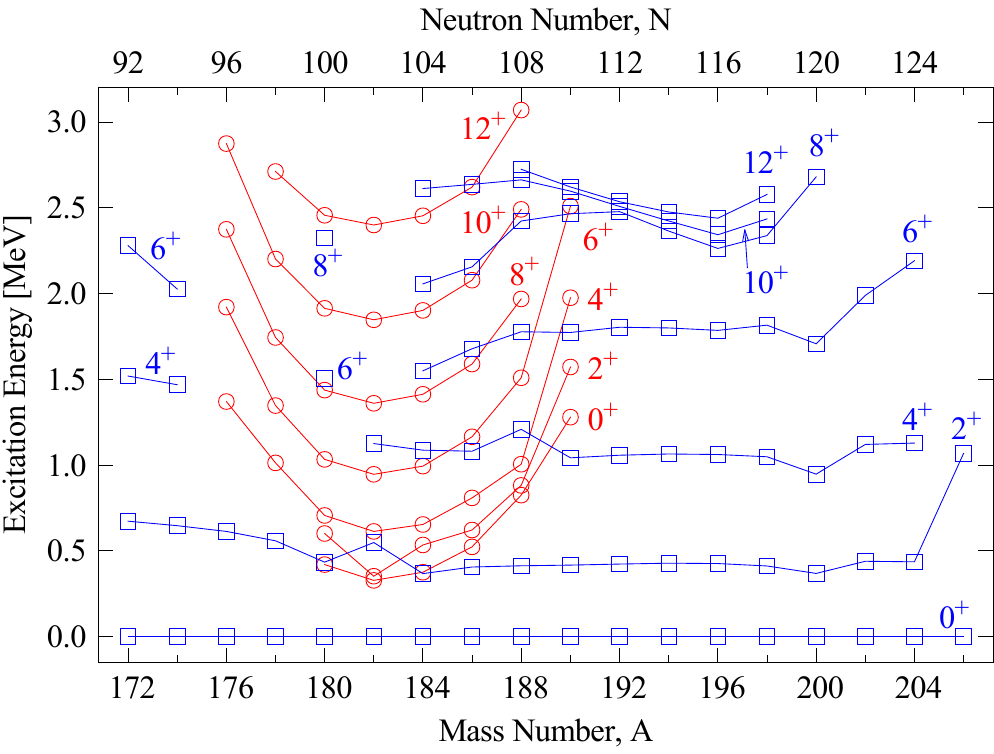}
\caption{\label{Hgsys}(Color online) Level energy systematics of even-mass mercury isotopes. Red circles refer to the assumed intruder states while blue squares refer to the assumed oblate states, guided by the results of the calculations in this work and that of Ref.~\cite{Richards1997}. The figure is an updated version of that in Ref.~\cite{Julin2001} using data taken from the NNDC database~\cite{NNDC}.}
\end{figure}

Spectroscopy of the even-mass mercury isotopes reveals a systematic trend of the intruding $0^{+}_{2}$ band head (shown in Fig. \ref{Hgsys}) which minimizes in energy near the neutron mid-shell at $N=104$. The excited states built upon these configurations become yrast above $I^{\pi}=4^{+}$ for $A\le186$, and the $2^{+}$ levels become close enough in energy to mix strongly.

% Hg level-energy systematics

Large $E0$ components in $2^{+}_{2}\rightarrow2^{+}_{1}$ transitions indicate a large degree of mixing. An attempt to understand the mixing between the bands was made by measuring this $E0$ component in-beam in $^{180}$Hg~\cite{Page2011} and $^{186}$Hg~\cite{Scheck2011}. The conversion coefficient of this transition can also be measured following $\beta$ decay, and there is an effort to provide more experimental data in this region~\cite{Elseviers2011}. Mixing of the $0^{+}$ states can be quantified by comparison of $\rho(E0)^{2}$ values~\cite{Kantele1979,Heyde1988} which have been experimentally determined in $^{180}$Hg~\cite{Wauters1994}, $^{184}$Hg~\cite{Cole1976}, and $^{188}$Hg~\cite{Joshi1994}.
% 'reproduce this behaviour', what behaviour?... REFEREE COMMENT --> observed mixing
As described in a recent review on the topic~\cite{Heyde2011}, the microscopic shell-model approach and the theoretical mean-field approach can both successfully reproduce the observed mixing. 
An analysis of $\alpha$-decay hindrance factors~\cite{Wauters1993,Wauters1994} indicates a smaller prolate contribution to the ground-state band $0^{+}$ state.

\begin{figure}[tb]
\includegraphics[width=\columnwidth]{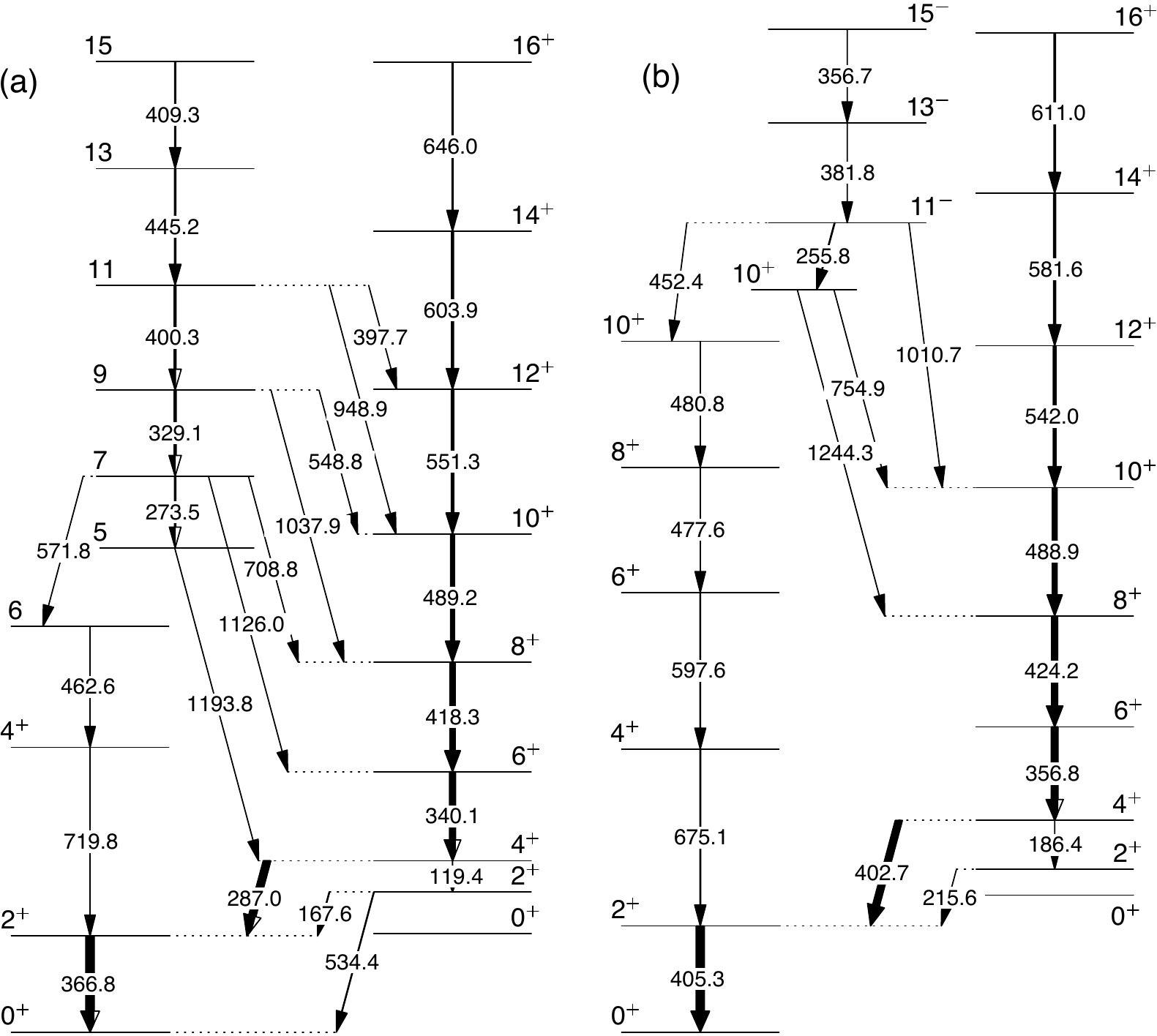}
%\subfloat[$^{184}$Hg]{\includegraphics[width=105pt]{184Hg_simple.pdf}}
%\subfloat[$^{186}$Hg]{\includegraphics[width=105pt]{186Hg_simple.pdf}}
\caption{\label{levelscheme}Partial level schemes of (a) $^{184}$Hg and (b) $^{186}$Hg showing states of interest. Data are taken from Refs.~\cite{Deng1995,Ma1993}.}
\end{figure}

To determine the magnitude and type of deformation of the two bands, their mixing strength, and to test the picture of shape coexistence, more precise data are required on the absolute transition strengths between the excited states of the nuclei in this mass region. Lifetimes of excited states in $^{180,182}$Hg have been recently measured by this collaboration~\cite{Grahn2009,Scheck2010}. To extend this knowledge and address these missing data, lifetime measurements of excited states in $^{184,186}$Hg have been performed. Partial level schemes of the nuclei studied in this work are given in Fig. \ref{levelscheme}.

% Level scheme - 184Hg & 186Hg

\section{Experimental Details}
Excited states in $^{184}$Hg and $^{186}$Hg were populated in the heavy-ion induced, fusion-evaporation reactions $^{148}$Sm($^{40}$Ar,$4n$)$^{184}$Hg and $^{150}$Sm($^{40}$Ar,$4n$)$^{186}$Hg, at beam energies of 200~MeV and 195~MeV. The average recoil velocity was $v/c=1.94\%$ and 1.90\%, respectively. 
The primary beam was provided by the ATLAS facility at the Argonne National Laboratory and delivered to the target position inside the Gammasphere spectrometer. The latter nominally consists of 110 Compton-suppressed, high-purity Ge detectors~\cite{Lee1990} arranged into 17 rings of constant polar angle, $\theta$, with respect to the beam. For this experiment, 100 detectors split into 16 rings were in use. The K\"oln plunger device was installed at the target position to allow for the Recoil Distance Doppler-Shift (RDDS) lifetime measurements. The distance of the 11~mg/cm$^{2}$-thick Au stopper foil was varied with respect to the 0.6~mg/cm$^{2}$ thick Sm target within a range of 2--2000~$\mu$m and data were taken at 12 distances for $^{184}$Hg (10 for $^{186}$Hg).

% Explain why 184 has more stats than 186... REFEREE COMMENT -> where measurement times were shorter
For the analysis of the data, $\gamma\gamma$-coincidence matrices were built using \texttt{GSSORT}~\cite{GSSORT} and the \texttt{ROOT} framework~\cite{Brun1997}. The data from different HPGe detectors of Gammasphere were grouped by rings at similar angles. In the analysis of $^{184}$Hg, the lifetimes are obtained by taking the weighted average of the lifetimes for different rings. Here, it was possible to obtain the velocity independently for each ring, utilizing the measured Doppler shift of the peaks. In the case of $^{186}$Hg, where measurement times were shorter, the lifetimes were derived directly from the weighted average of the intensities whilst the velocity is obtained by weighting the velocities from the individual $\gamma\gamma$-coincidence spectra for all combinations of ring pairs.

% 184Hg shifts vs. distance plot
\begin{figure}[tb]
\includegraphics[width=1.0\columnwidth]{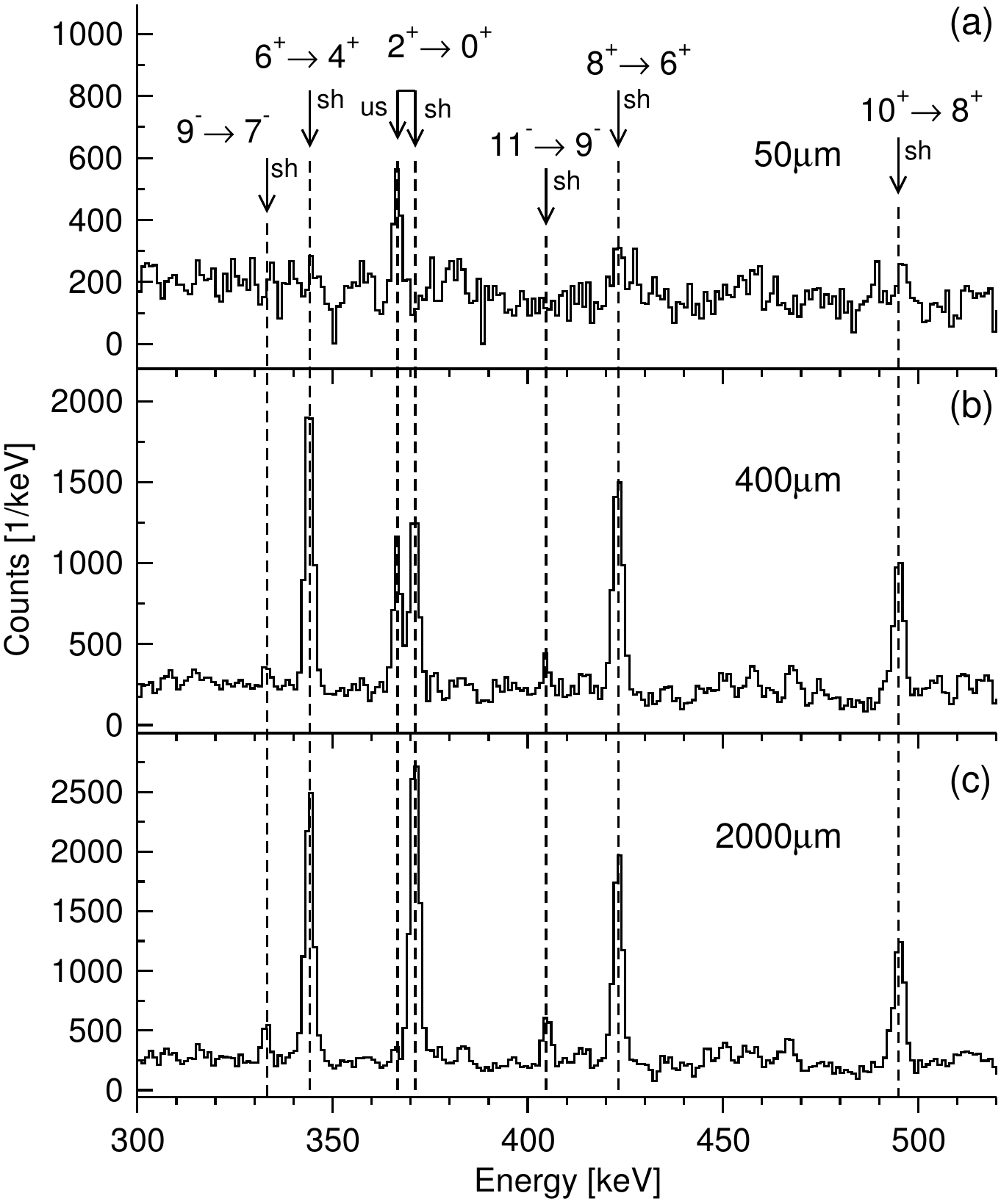}
\caption{\label{spec}Gamma-ray spectra from the Gammasphere detectors at $\theta\approx 53^{\circ}$, gated on the shifted (sh) component of the $4_{1}^{+}\rightarrow 2_{1}^{+}$ transition in $^{184}$Hg at a target-to-stopper distance of (a)~50~$\mu$m, (b)~400~$\mu$m, and (c)~2000~$\mu$m. Transitions which feed the $4_{1}^{+}$ state only have their shifted (sh) component in coincidence, while one also observes the unshifted (us) component of the $2_{1}^{+}\rightarrow 0_{1}^{+}$ depopulating transition.}
\end{figure}

Typical $\gamma\gamma$-coincidence spectra obtained during the experiment are presented in Fig.~\ref{spec}, in which it is possible to see the variation of intensity of the fully Doppler-shifted component of the depopulating transition, $I_{\mathrm{sh}}$, with increasing distance. At the smallest distances the flight times of the nuclei are very short leading to few $\gamma$ rays emitted in-flight. Consequently, when gating in the $\gamma$-$\gamma$ matrix, a low number of events are observed, an effect present in Fig.~\ref{spec}~(a).

To account for the varying measurement time at different target-to-stopper distances, the intensities had to be normalized. For $^{184}$Hg, the distances were normalized by employing a gate on the total intensity of the $340$-keV, $6_{1}^{+}\rightarrow4_{1}^{+}$ transition and averaging the intensities of the $287$-keV, $4_{1}^{+}\rightarrow 2_{1}^{+}$ and $367$-keV, $2_{1}^{+}\rightarrow 0_{1}^{+}$ transitions at all angles. For $^{186}$Hg, all possible $\gamma\gamma$ coincidences for the yrast transitions up to the $10_{1}^{+}$ state, in spectra with $\theta<90^{\circ}$, were used for the normalization. The use of the coincidence differential decay curve method (DDCM) eradicates the influence of recoil de-orientation~\cite{Petkov1994}.

In the analysis of $^{184}$Hg, lifetimes were obtained using the prescriptive coincidence DDCM~\cite{Bohm1993,Dewald2012}. 
% Must explain this more here... REFEREE COMMENT
While here we explain only the features of the method that are required for this analysis, the full details are contained in Refs.~\cite{Bohm1993,Dewald2012}, to which the reader is referred for more information.
The coincidence technique allows for the elimination of systematic errors usually introduced in RDDS-singles measurements by the unknown feeding history from states that lie higher in energy.
%Energy gates are placed on the fully-shifted component of the transition which directly feeds the state of interest.
The decay curves, as a function of distance, $x$, for this analysis were constructed using gates on the shifted component of the yrast transition directly feeding the state of interest, ensuring a significant simplification whereby the lifetime is determined using only the ratio of the unshifted ($I_{\mathrm{us}}$) and the time derivative %($\frac{d}{dt}$)
of the fully-shifted ($I_{\mathrm{sh}}$) components of the depopulating transition.
\begin{equation} \label{eq:tau}
\tau(x) = \frac{ I_{\mathrm{us}}(x) }{ \frac{d}{dt} I_{\mathrm{sh}}(x) }	,
\end{equation}

\begin{figure}[tb]
\includegraphics[width=\columnwidth]{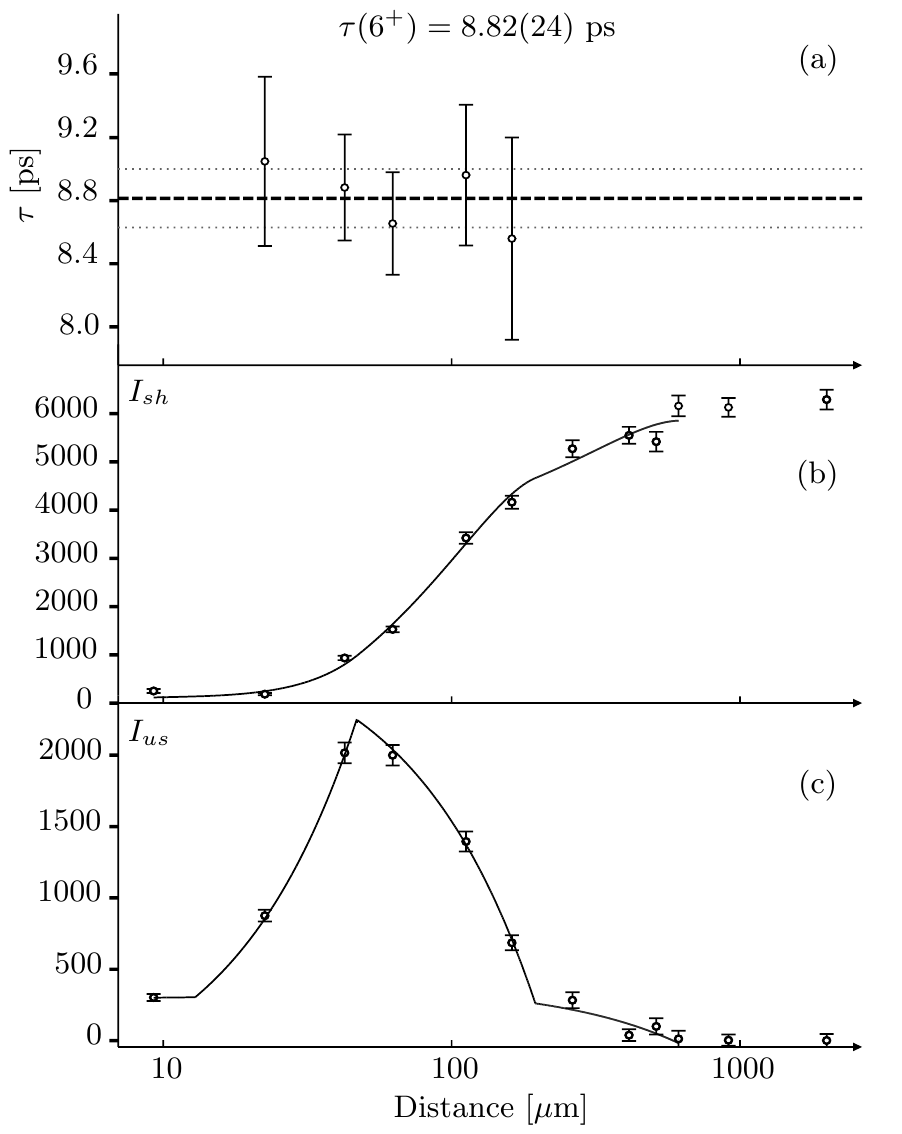}
\caption{(a) Lifetime values, $\tau$, extracted at each distance in the sensitive region, for the 340-keV, $6_{1}^{+}\rightarrow4_{1}^{+}$ transition in $^{184}$Hg, measured in the detectors at $\theta\approx 53^{\circ}$. The dashed and dotted lines show the weighted average and the associated uncertainty, respectively. (b) Normalized intensity curve for the shifted component, $I_{\mathrm{sh}}$, fitted (solid line) with a function consisting of continuous, piecewise second-order polynomials. (c) Normalized intensity curve for the unshifted component, $I_{\mathrm{us}}$. The curve in (c) is proportional to the derivative of that in (b) and both curves are fitted simultaneously. According to Eq.~\ref{eq:tau}, the ratio of these gives the lifetime of the state, shown in (a).}
\label{ddcm}
\end{figure}

Typical fits of continuously-connected second-order polynomials, performed with \texttt{Napatau}~\cite{Napatau,Dewald1989}, are illustrated in Fig. \ref{ddcm}. The lifetime is determined at every distance and should sit at a constant value. Deviations from this behavior indicate systematic effects, which can be identified easily with this method. The weighted average, $\tau_{av}$, is taken of the points inside of the sensitivity region, i.e. where the derivative of the decay curve is largest.

% DDCM plots

\begin{figure}[tb]
\centering
\includegraphics[width=\columnwidth]{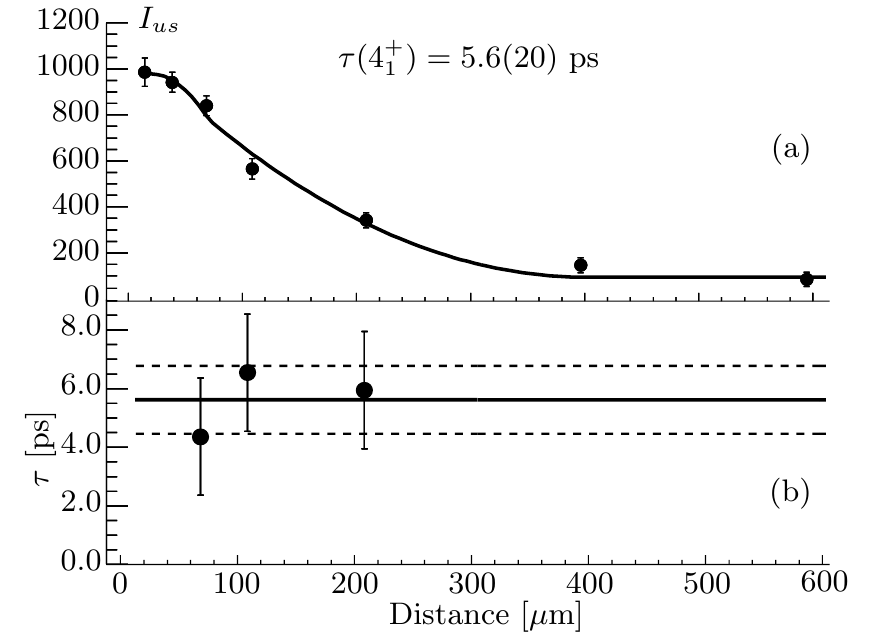}
\caption{(a) The normalized intensity of the unshifted component of the $4_{1}^{+} \rightarrow 2_{1}^{+}$ depopulating transition in $^{186}$Hg. The fitted line is as in Fig.~\ref{ddcm}. (b) Lifetime values extracted at each distance in the sensitive region. The solid and dashed lines show the weighted average and the associated uncertainty.}
\label{Fig:4plusTauPlot}
\end{figure}

Lifetimes of the yrast states of ${}^{186}$Hg up to the $10_{1}^{+}$ state were similarly determined with the exception of the $4_{1}^{+}$ state. In this case, the near doublet of the $4_{1}^{+} \rightarrow 2_{1}^{+}$ and $2_{1}^{+} \rightarrow 0_{1}^{+}$ transitions rendered it problematic to use the simple ``gate from above'' method. Instead, the method of ``gating from below''~\cite{Petkov2001} was used. The corresponding $\tau$ plot for the $4_{1}^{+}$ state is found in Fig.~\ref{Fig:4plusTauPlot}. The intensities involving the $6_{1}^{+}$ and $4_{1}^{+}$ analysis were corrected for a contamination from the $15_{2}^{-} \rightarrow 13_{2}^{-}$ transition, using a gate on the $13_{2}^{-} \rightarrow 11_{2}^{-}$ transition.

\section{Results}

The final weighted averages of the mean lifetimes of all states studied are shown in Table \ref{results} along with the transition strengths, $B(E2)$, and absolute transitional quadrupole moments, $|Q_{t}|$, of the depopulating yrast transitions.
Transition quadrupole moments, $Q_{t}$, are related to the $B(E2)$ values assuming a rotating quadrupole deformed nucleus using the rotational model:
\begin{equation}
B(E2; I \rightarrow I^{\prime}) = \frac{5}{16\pi} {\langle I 0 2 0 | I^{\prime}  0 \rangle}^{2} Q_{t}^{2}	,
\end{equation}

\noindent where $\langle I 0 2 0 | I^{\prime}  0 \rangle$ is a Clebsch-Gordan coefficient and ${I^{\prime}=I-2}$.

\begin{figure}[tb]
\includegraphics[width=\columnwidth]{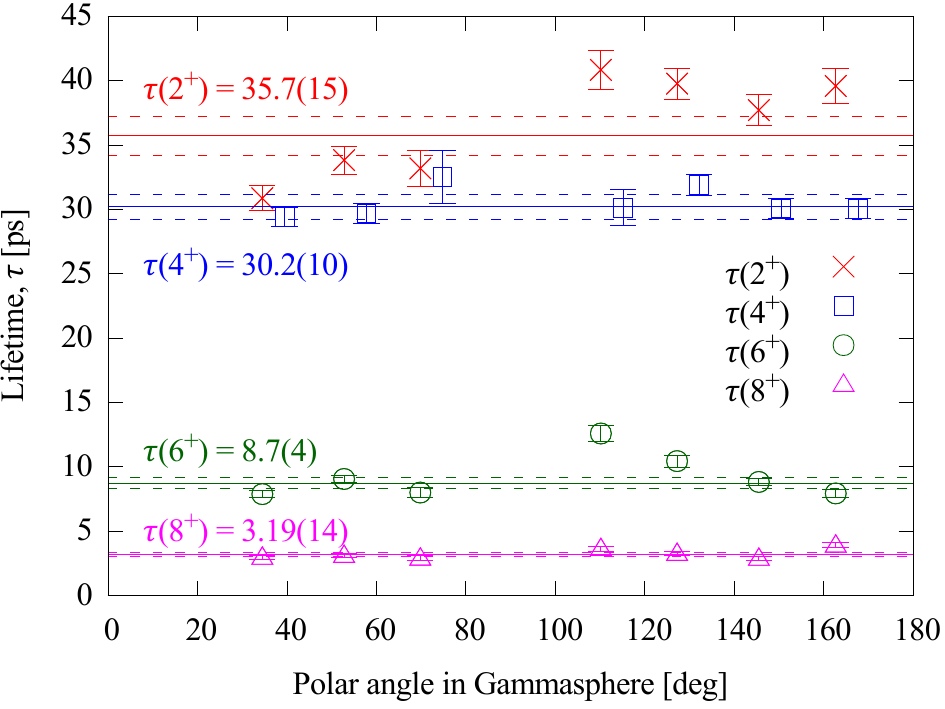}
\caption{\label{fig:lifetime184Hg}(Color online) Lifetimes of yrast states in $^{184}$Hg as a function of angle. The weighted average and 1$\sigma$ uncertainty, calculated from the weighted standard deviation, for each state are denoted by the solid and dashed lines, respectively. Please note, the $\tau(4^{+})$ values are offset by $5^{\circ}$ on the x-axis to maintain clarity and some individual error bars may be smaller than the marker size.}
\end{figure}

In $^{184}$Hg, the seven independent measurements of each of the lifetimes of the even-spin yrast states are presented in Fig.~\ref{fig:lifetime184Hg}, as a function of the ``ring'' angle at which they were determined in Gammasphere.

It is worth noting that the new lifetime for $4^{+}_{1}$ state in $^{186}$Hg is smaller than the previously measured value from Ref.~\cite{Proetel1974}. The discrepancy in the first measurement is likely due to complications that the authors encountered in resolving the doublet and subsequent assumptions which were made.

% Lifetimes in 184Hg vs angle

% Results table
\begin{table*}[tb]
\caption{\label{results}Properties of the states investigated in this study. The uncertainties presented on $\tau_{\mathrm{av}}$ represent the $1\sigma$ statistical error and include an additional systematic uncertainty which accounts for the choice of the fitting function and relativistic effects~\cite{Dewald2012}, typically $\le3\%$. Gamma-ray energies~($E_{\gamma}$) and branching fractions~(b.f.) of the depopulating $\gamma$-ray transitions (corrected for internal conversion), as well as spin and parity~($I^{\pi}$) values, are taken from Refs.~\cite{Deng1995, Ma1993}. In cases where only one depopulating transition is observed b.f. is assumed to be equal to unity. $\tau_{\mathrm{prev}}$ values from Refs.~\cite{Rud1973,Ma1986,Proetel1974} are shown for comparison.}
\begin{ruledtabular}
\begin{tabular}{p{8mm}cccccccccccc}
	& $I^{\pi}$~($\hbar$) & $E_{\gamma}$ (keV) & b.f. & $\tau_{\mathrm{av}}$  (ps) & $\tau_{\mathrm{prev}}$  (ps) & $B(E2)\downarrow$ (W.u.) & $|Q_{t}|$~($e$b) \\[2pt]
\hline
$^{184}$Hg & $2_{1}^{+}$ 	& 366.8	& 1			& 35.7(15) & 30(7)	  			& 52(2)	& 4.04(8) \\
 		   & $4_{1}^{+}$ 	& 287.0	& 0.959(4)		& 30.2(10)	 & 32.8(34)			& 191(6)	& 6.46(11) \\
		   & $6_{1}^{+}$ 	& 340.1	& 1			& 8.7(4)	 & 8.1(31)				& 308(15) & 7.81(19) \\
		   & $8_{1}^{+}$ 	& 418.3	& 1	 		& 3.19(14)	 & $2.9^{+1.1}_{-1.6}$	& 309(13) & 7.65(17) \\
		   & $9_{3}$  	& 329.1	& 0.65(16)		& 12.1(8)	 & $ - $				& 169(40) & 5.6(7) \\[4pt]
		   
$^{186}$Hg & $2_{1}^{+}$		& 405.3	& 1			& 24(3)	 & 26(4)		& 47(6)	& 3.9(2) \\
 		   & $4_{1}^{+}$		& 402.7	& 0.93(2)		& 5.6(20)	 & 13(4)		& 200(70)	& 6.6(12) \\
		   & $6_{1}^{+}$		& 356.8	& 1			& 9.1(4)	 & 7(3)		& 231(10)	& 6.82(15) \\
		   & $8_{1}^{+}$ 	& 424.2	& 1	 		& 4.5(3)	 & $\approx 4$	& 202(14)	& 6.2(2) \\
		   & $10_{1}^{+}$	& 488.9	& 1			& 1.9(2)	 & $ - $		& 238(25)	& 6.7(4) \\
\end{tabular}
\end{ruledtabular}
\end{table*}

\section{Discussion}

\subsection{Yrast states in $^{184}$Hg and $^{186}$Hg}
\begin{figure}[tb]
\includegraphics[width=\columnwidth]{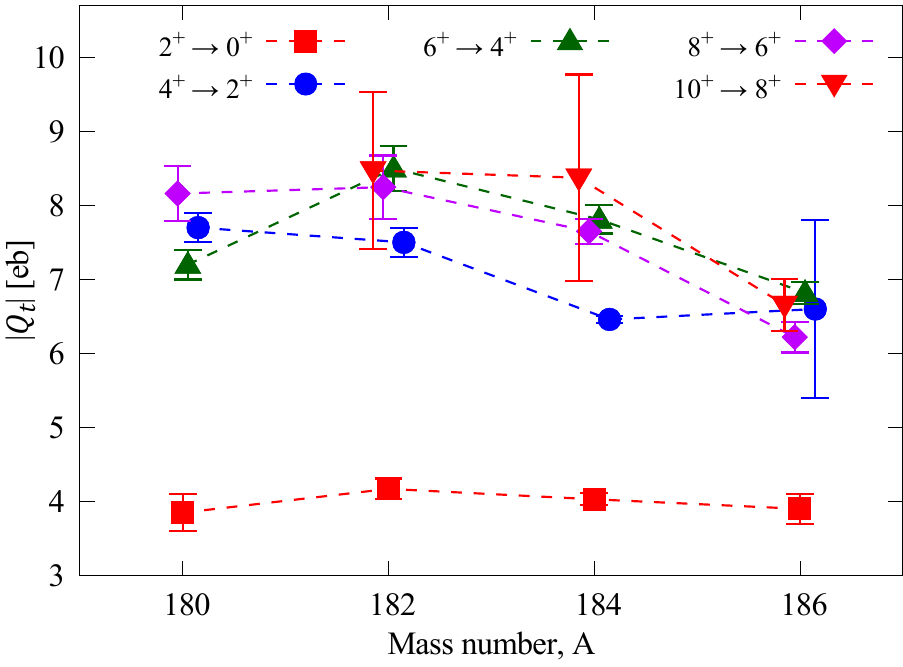}
\caption{(Color online) Experimental $|Q_{t}|$ values, extracted from measured lifetimes, for yrast transitions in mercury nuclei as a function of mass number, $A$. Data for $A=180,182$ are taken from Ref.~\cite{Grahn2009} and the $10^{+}\rightarrow 8^{+}$ value for $A=184$ is taken from Ref~\cite{Ma1986}. Some markers are slightly offset from integer $A$ values to maintain clarity.}
\label{Qtsys}
\end{figure}

Transitional quadrupole moments, $|Q_{t}|$, for the even-spin yrast states are given in Fig. \ref{Qtsys} for the mercury isotopes where $180 \le A \le 186$. The $2^+$ states show no strong variation in $|Q_{t}|$ with mass number, whereas the collectivity of the $4^{+}$ states reduces with increasing mass number. This can be compared to the energy level systematics in Fig.~\ref{Hgsys} where the energy of the intruder states reaches a minimum at $A=182$.

%
%\subsection{The $9_{3}^{\left(-\right)}$ state in $^{184}$Hg}
\subsection{The $9_{3}$ state in $^{184}$Hg} % drop the (-) according to REFEREE COMMENT
After the even-spin, positive-parity yrast band in $^{184}$Hg, the most populated band is the odd-spin rotational band built upon the $I=5$ state, observed at 1.848~MeV, which becomes yrast at around 4~MeV. 
Analogous bands have been observed in the neighboring isotopes, specifically $^{178}$Hg~\cite{Kondev1999} and $^{180}$Hg~\cite{Kondev2000} and their structure discussed in terms of octupole correlations. Lifetime measurements of states in this band in $^{182}$Hg have been performed and a consistency in the structure of these states has been extended to $^{184}$Hg using the energy displacement of states differing by $3\hbar$~\cite{Scheck2010}. The quadrupole moment measured here for the $9_{3}\rightarrow7_{3}$ transition in $^{184}$Hg, $|Q_{t}|=5.6(7)$~$e$b, is similar to that of the even-spin yrast band, $|Q_{t}|\simeq7.7$~$e$b, although it is smaller than those measured in the lighter isotopes. 

% Qt systematics in Hg

\subsection{Two-state mixing calculations}
In order to shed light on the properties of the coexisting structures in these light mercury isotopes, phenomenological two-band mixing calculations have been carried out using the assumption of a spin-independent interaction between two rotational structures. In the calculations, the variable moment of inertia (VMI) model~\cite{Mariscotti1969} was used to fit known level energies of rotational bands built upon the first two $0^{+}$ states, up to and including $I^{\pi}=10^{+}$ and $4^{+}$ for yrast and non-yrast bands, respectively. Employing the method of Lane~\textit{et al.}~\cite{Lane1995}, one can derive the wave-function amplitudes of the two configurations from the mixing strengths. These are shown in Table~\ref{tab:mixing} along with the mixing strength, $V$, for each isotope and the band-head energies of the two configurations.

% Mixing amplitudes etc.
\begin{table}[tb]
\caption{Wave-function amplitudes of the normal configuration, $\alpha_{I}$, at each spin, $I$, and spin-independent interaction strengths between members of the normal ($n$) band and intruder ($i$) band, $V=|\langle I_{n} | V_{I} | I_{i} \rangle|$ (equal for all values of $I$), calculated using the model described in the text. The extracted unperturbed band-head energies are denoted by $E^{n}_{0}$ and $E^{i}_{0}$ for the normal and intruder bands, respectively.} \label{tab:mixing}
\begin{ruledtabular}
\begin{tabular}{cccc}
Nucleus &  & $I$~[$\hbar$] & $|\alpha_{I}|$ \\[2pt]
\hline
\multirow{7}{*}{$^{180}$Hg}
& \multirow{2}{*}{$V = 82.1$~keV}				& 	0   & 0.9799 \\
&										&	2   & 0.7722 \\
& \multirow{2}{*}{$E^{n}_{0}=16.7$~keV}			&	4   & 0.1571 \\
& 										&	6   & 0.0817 \\
& \multirow{2}{*}{$E^{i}_{0}=403.3$~keV}			&	8   & 0.0546 \\
& 										&	10 & 0.0408 \\[2pt]
\hline
\multirow{7}{*}{$^{182}$Hg}
& \multirow{2}{*}{$V = 89.4$~keV}				&	0   & 0.9606 \\
&										&	2   & 0.5382 \\
& \multirow{2}{*}{$E^{n}_{0}=25.9$~keV}			&	4   & 0.1781 \\
& 										&	6   & 0.1003 \\
& \multirow{2}{*}{$E^{i}_{0}=309.0$~keV}			&	8   & 0.0697 \\
& 										&	10 & 0.0534 \\[2pt]
\hline
\multirow{7}{*}{$^{184}$Hg}
& \multirow{2}{*}{$V = 84.7$~keV}				&	0   & 0.9725 \\
&										&	2   & 0.7172 \\
& \multirow{2}{*}{$E^{n}_{0}=20.4$~keV}			&	4   & 0.2014 \\
& 										&	6   & 0.1025 \\
& \multirow{2}{*}{$E^{i}_{0}=353.4$~keV}			&	8   & 0.0679 \\
& 										&	10 & 0.0506 \\[2pt]
\hline
\multirow{7}{*}{$^{186}$Hg}
& \multirow{2}{*}{$V = 66.3$~keV}				&	0   & 0.9915 \\
&										&	2   & 0.9506 \\
& \multirow{2}{*}{$E^{n}_{0}=8.9$~keV}			&	4   & 0.2604 \\
& 										&	6   & 0.0978 \\
& \multirow{2}{*}{$E^{i}_{0}=505.4$~keV}			&	8   & 0.0587 \\
& 										&	10 & 0.0416 \\[2pt]
\hline
\multirow{7}{*}{$^{188}$Hg}
& \multirow{2}{*}{$V = 76.3$~keV}				&	0   & 0.9954 \\
&										&	2   & 0.9884 \\
& \multirow{2}{*}{$E^{n}_{0}=7.8$~keV}			&	4   & 0.8933 \\
& 										&	6   & 0.2467 \\
& \multirow{2}{*}{$E^{i}_{0}=791.8$~keV}			&	8   & 0.1095 \\
& 										&	10 & 0.0691 \\[1pt]
\end{tabular}
\end{ruledtabular}
\end{table}

\begin{figure}[tb]
\includegraphics[width=\columnwidth]{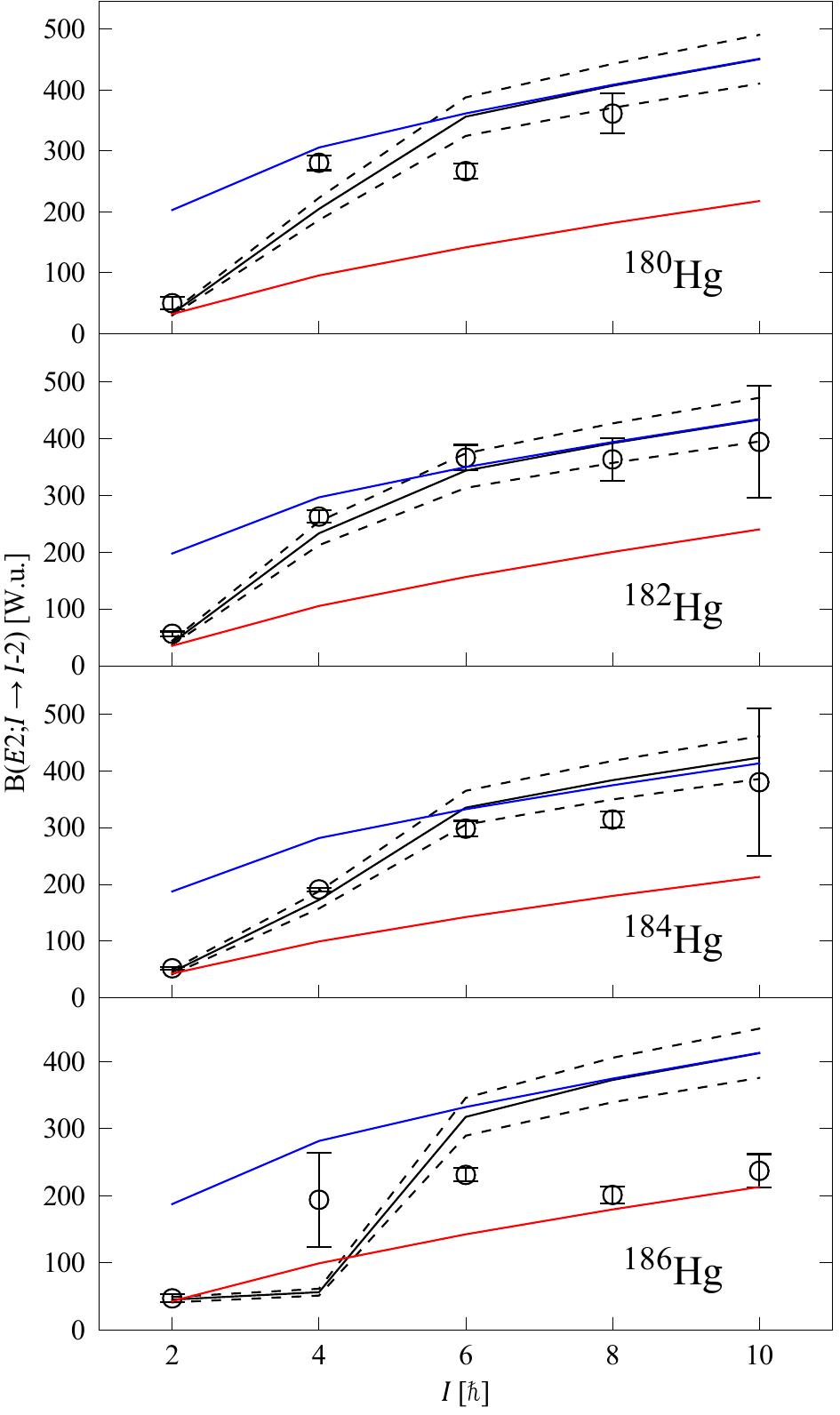}
\caption{\label{BE2_mixfig}(Color online) Experimental $B(E2)$ values measured in this work (points) plotted as a function of spin and compared to those extracted from the mixing calculations (solid black line, dashed line represents the uncertainty in the constant, $k$).
For reference, the intra-band $B(E2)$ values calculated for the pure unperturbed normal (red) and intruder (blue) bands are also shown.
Data for $^{180,182}$Hg are taken from Ref.~\cite{Grahn2009} and the point at $I=10$~$\hbar$ in $^{184}$Hg is taken from Ref.~\cite{Ma1986}.}
\end{figure}

Similar mixing calculations have previously been performed for $^{180,182,184}$Hg~\cite{Dracoulis1994}.
The present results for $^{180,182}$Hg differ due to the inclusion of non-yrast states identified in recent
studies~\cite{Elseviers2011,Page2011}, which provide additional constraints on the calculations.
The calculation for $^{184}$Hg presented in Table~\ref{tab:mixing} is essentially identical to that in Ref.~\cite{Dracoulis1994}.
It places the $I=2$ member of the intruder band just 5~keV above the corresponding state in the normal band before mixing.
This leads to almost complete mixing between the states, producing a first-excited $2^{+}$ state which comprises 51\% of the normal configuration and 49\% of the intruder configuration.
An alternative calculation in which the order of these states is reversed was also performed and yielded very similar parameters to those presented in Table~\ref{tab:mixing}.
In this scenario the unmixed states are nearly degenerate, so the degree of mixing is even greater.
The resulting $B(E2)$ values (presented later in Fig.~\ref{BE2_mixfig}) are not significantly different, so in what follows only the results from the former calculation will be discussed.

Low-lying levels in $^{184,186}$Hg have also been interpreted assuming the mixing of spherical and deformed states~\cite{Dickmann1974}, while an alpha-plus-rotor model has been used to extract spin-dependent interaction strengths and mixing amplitudes in $^{182,184,186}$Hg~\cite{Richards1997}. The latter study predicted a contribution of the more-strongly deformed structure to the $2_{1}^{+}$ state in $^{182}$Hg of 76\%, comparing very well to the value of 71\% obtained in this work. We note here that this contribution drops to only 2.3\% for the same state in $^{188}$Hg.
The corresponding isotones in the platinum nuclei were also interpreted recently using two-band mixing calculations and qualitatively similar conclusions were drawn regarding a strong degree of mixing for the low-spin states~\cite{Dracoulis1994,Walpe2012}.

It is possible to determine the transitional quadrupole moment of the unperturbed $I \rightarrow I-2$ transitions in the normal ($n$) and intruder bands ($i$), $Q_{I}$, using an average of the moment of inertia of the two states, ${\mathcal J}^{\prime}_{I}$, such that~\cite{Mariscotti1969}
\begin{equation}
Q_{I} = k \sqrt{{\mathcal J}^{\prime}_{I}}= k \sqrt{\left({\mathcal J}_{I}+{\mathcal J}_{I-2}\right)/2} ,
\end{equation}

\noindent where ${\mathcal J}_{I}$ is the moment of inertia for a pure state with spin $I$, extracted from the fit. An evaluated value of the constant, fitted to data in the neutron-deficient $A=170$ region, $k=45(2)$~$e$b~keV$^{-1/2}$~\cite{Dracoulis1988}, was used in this study. Combining knowledge of the wave-function amplitudes with the intrinsic quadrupole moments of the pure states, $Q_{I}$, it is possible to extract the $B(E2; I \rightarrow I-2)$ values of the mixed states~\cite{Lane1995}.
The relative sign of the intrinsic quadrupole moments of the two configurations must be assumed to be positive or negative and was found to be best reproduce the data when positive. 
This feature has been noted in previous calculations~\cite{Guttormsen1981,Lane1995} and is at odds with what is expected from the rotational model when two bands with the same $K$ quantum number have an opposite sign of deformation, i.e. an oblate and a prolate band.

The results of the calculations for the yrast sequences are compared in Fig.~\ref{BE2_mixfig} with those extracted from the lifetimes measured in this work. Good agreement is found for the majority of yrast transitions, even for cases where the states are strongly mixed. The observed discrepancies for $I\ge6$ in $^{186}$Hg may indicate a breakdown of the simple two-band picture as other structures begin to influence the yrast states~\cite{Helppi1983,Delaroche1994}. For the nuclei presented in Fig.~\ref{BE2_mixfig}, the experimental and calculated $B(E2; 2_{1}^{+}\rightarrow0_{1}^{+})$ values are similar in each isotope, while the $B(E2; 4_{1}^{+}\rightarrow2_{1}^{+})$ varies and, in each case, is not consistent with a transition within either of the pure bands. 
The low $B(E2; 2_{1}^{+} \rightarrow 0_{1}^{+})$ value is interpreted as being due to a transition between the weakly-deformed oblate $0^{+}$ ground state and a more strongly-deformed prolate $2^{+}$ state~\cite{Grahn2009}. However, the admixture of the normal and intruder configurations for the $0^{+}$ and $2^{+}$ states is unique in each isotope and a similar $B(E2)$ value is not necessarily indicative of similar structures across the mass range.

% BE2 vs spin for experiment and mixing model

\begin{figure}[tb]
\includegraphics[width=\columnwidth]{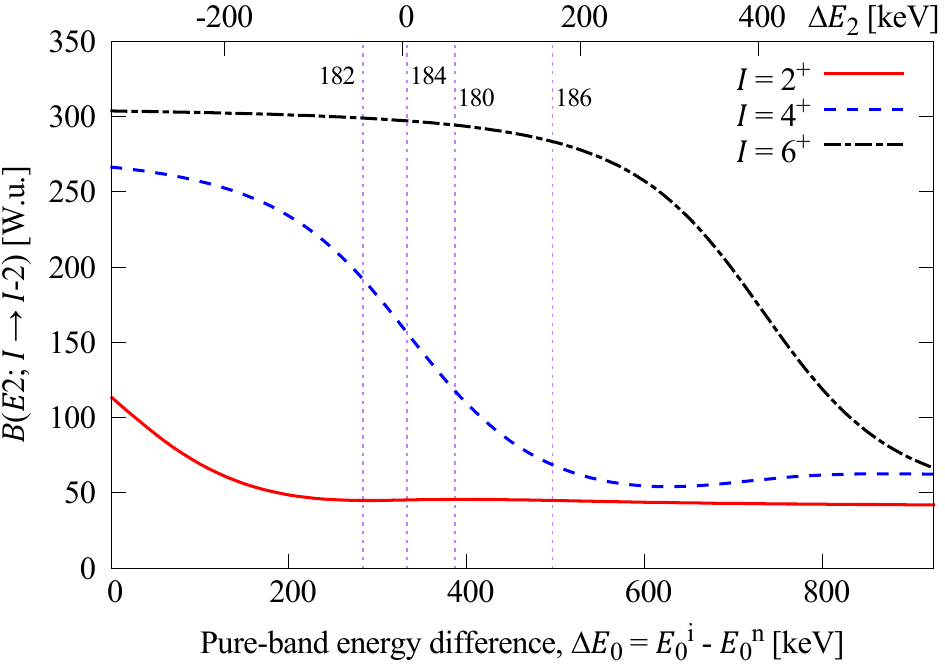}
\caption{\label{be2_deltaEpure}(Color online) Representative $B(E2; I_{1}^{+} \rightarrow (I-2)_{1}^{+})$ values as a function of the energy separation of the pure band-heads, $\Delta E_{0}$. The curves for $I=2,4,6$ are calculated assuming the mixing strength and moments of inertia are equal to that extracted for $^{184}$Hg. As a guide, the energy difference between the pure $2^{+}$ states, $\Delta E_{2}$, under this assumption is also shown. The dashed vertical lines show the extracted $\Delta E_{0}$ values for each of the four isotopes.}
\end{figure}

In Fig.~\ref{be2_deltaEpure}, calculated $B(E2)$ values are plotted as a function of the energy difference between the pure, unmixed $0^{+}$ band heads. The parameters used in the calculations are those of $^{184}$Hg, but since they are not too dissimilar for all of the isotopes, the curves can be considered representative of the mercury isotopes around the neutron mid-shell. The four isotopes compared have $\Delta E_{0}$ values in the range of 250--550~keV, marked by the vertical lines in Fig.~\ref{be2_deltaEpure}. One clearly observes that above 200~keV the $B(E2;2_{1}^{+}\rightarrow0_{1}^{+})$ values remain constant, even though the square of the mixing amplitude of the $I=2$ states, $\alpha_{2}^{2}$, drops from 0.80 at 200 keV down to 0.015 at 1 MeV. In contrast, the $B(E2;4_{1}^{+}\rightarrow2_{1}^{+})$ values vary significantly in the range of interest and are much more sensitive to $\alpha_{2}$, while the $B(E2;6_{1}^{+}\rightarrow4_{1}^{+})$ values become more sensitive at larger band-head energy differences.

% BE2 vs pure band-head energy difference

\section{Summary and conclusions}

Lifetimes of excited states have been measured by employing the RDDS technique. Yrast states up to $I^{\pi}=8^{+}$ in $^{184}$Hg and $I^{\pi}=10^{+}$ in $^{186}$Hg have been studied. Distinct differences in deformation for the assigned normal and intruder bands could be shown. The lifetime of the $4_{1}^{+}$ state in $^{186}$Hg was found to be shorter than the previously measured value. Lifetimes of the $9_{3}$ state in $^{184}$Hg and the $10^{+}$ state in $^{186}$Hg have been measured for the first time. The other lifetimes are consistent with previous measurements, while the uncertainty could be reduced significantly. These more precise lifetime values, with those of Ref.~\cite{Grahn2009}, have been a vital input to the analysis of Coulomb excitation experiments at the REX-ISOLDE facility~\cite{Petts2009,Bree2013}.

Rotational bands built upon the first two $I^{\pi}=0^{+}$ states have been considered in terms of a two-state mixing model and the mixing amplitudes of the two configurations extracted as a function of spin. It is observed that, while the ground state remains composed of predominantly one configuration, namely the assumed weakly-deformed normal structure, in all of the even-mass mercury isotopes considered ($180 \le A \le 188$), the first excited $2^{+}$ state changes dramatically in its composition. This is in contrast to a na\"{\i}ve interpretation emanating from the systematics of the $2^{+}$ level energies and $B(E2; 2^{+} \rightarrow 0^{+})$ values, which are both strikingly similar across the mass range. 

\begin{acknowledgments}
The authors would like to thank the operators of the ATLAS facility at ANL for providing the beams.
This work was supported by the United Kingdom Science and Technology Facilities Council, by the German DFG, grant No. DE 1516/1-1, by the Academy of Finland, contract number 131665, by the German BMBF, grant No. 06 KY 7153, by US Department of Energy, Office of Nuclear Physics, under Contract No. DE-AC02-06CH11357 as well as Grant No. DE-FG02-94ER40834, by FWO-Vlaanderen (Belgium), by GOA/2010/010 (BOF KU Leuven), by the Interuniversity Attraction Poles Programme initiated by the Belgian Science Policy Office (BriX network P7/12), by the European Commission within the Seventh Framework Programme through I3-ENSAR (contract no. RII3-CT-2010-262010) and by a Marie Curie Intra-European Fellowship of the European Community's 7th Framework Programme under contract number (PIEF-GA-2008-219175). L.P.G. acknowledges support from FWO-Vlaanderen (Belgium) as an FWO Pegasus Marie Curie Fellow.
\end{acknowledgments}

\bibliographystyle{aip_doi}
\bibliography{Mercury}

\begin{thebibliography}{10}

\bibitem{Bonn1972}
J.~Bonn, G.~Huber, H.-J. Kluge, L.~Kugler, and E.~Otten,
\newblock \href {http://dx.doi.org/10.1016/0370-2693(72)90253-5} {Physics
  Letters B {\bf 38}, 308 (1972)}.

\bibitem{Frauendorf1975}
S.~Frauendorf and V.~Pashkkevich,
\newblock \href {http://dx.doi.org/10.1016/0370-2693(75)90360-3} {Physics
  Letters B {\bf 55}, 365 (1975)}.

\bibitem{Ulm1986}
G.~Ulm et~al.,
\newblock \href {http://dx.doi.org/10.1007/BF01294605} {Zeitschrift f\"{u}r
  Physik A Atomic Nuclei {\bf 325}, 247 (1986)}.

\bibitem{Nazarewicz1993}
W.~Nazarewicz,
\newblock \href {http://dx.doi.org/10.1016/0370-2693(93)90107-S} {Physics
  Letters B {\bf 305}, 195 (1993)}.

\bibitem{Richards1997}
J.~D. Richards, T.~Berggren, C.~R. Bingham, W.~Nazarewicz, and J.~Wauters,
\newblock \href {http://dx.doi.org/10.1103/PhysRevC.56.1389} {Physical Review C
  {\bf 56}, 1389 (1997)}.

\bibitem{Julin2001}
R.~Julin, K.~Helariutta, and M.~Muikku,
\newblock \href {http://dx.doi.org/10.1088/0954-3899/27/7/201} {Journal of
  Physics G: Nuclear and Particle Physics {\bf 27}, R109 (2001)}.

\bibitem{NNDC}
J.~Tuli,
\newblock \href {http://www.nndc.bnl.gov/ensdf} {{Evaluated Nuclear Structure
  Data File (ENSDF)}}, 2012.

\bibitem{Page2011}
R.~D. Page et~al.,
\newblock \href {http://dx.doi.org/10.1103/PhysRevC.84.034308} {Physical Review
  C {\bf 84}, {034308} (2011)}.

\bibitem{Scheck2011}
M.~Scheck et~al.,
\newblock \href {http://dx.doi.org/10.1103/PhysRevC.83.037303} {Physical Review
  C {\bf 83}, {037303} (2011)}.

\bibitem{Elseviers2011}
J.~Elseviers et~al.,
\newblock \href {http://dx.doi.org/10.1103/PhysRevC.84.034307} {Physical Review
  C {\bf 84}, {034307} (2011)}.

\bibitem{Kantele1979}
J.~Kantele et~al.,
\newblock \href {http://dx.doi.org/10.1007/BF01435933} {Zeitschrift f\"{u}r
  Physik A Atoms and Nuclei {\bf 289}, 157 (1979)}.

\bibitem{Heyde1988}
K.~Heyde and R.~A. Meyer,
\newblock \href {http://dx.doi.org/10.1103/PhysRevC.37.2170} {Physical Review C
  {\bf 37}, 2170 (1988)}.

\bibitem{Wauters1994}
J.~Wauters et~al.,
\newblock \href {http://dx.doi.org/10.1103/PhysRevC.50.2768} {Physical Review C
  {\bf 50}, 2768 (1994)}.

\bibitem{Cole1976}
J.~Cole et~al.,
\newblock \href {http://dx.doi.org/10.1103/PhysRevLett.37.1185} {Physical
  Review Letters {\bf 37}, 1185 (1976)}.

\bibitem{Joshi1994}
P.~Joshi et~al.,
\newblock \href {http://dx.doi.org/10.1142/S021830139400019X} {International
  Journal of Modern Physics E {\bf 3}, 757 (1994)}.

\bibitem{Heyde2011}
K.~Heyde and J.~L. Wood,
\newblock \href {http://dx.doi.org/10.1103/RevModPhys.83.1467} {Reviews of
  Modern Physics {\bf 83}, 1467 (2011)}.

\bibitem{Wauters1993}
J.~Wauters et~al.,
\newblock \href {http://dx.doi.org/10.1007/BF01290335} {Zeitschrift f\"{u}r
  Physik A Hadrons and Nuclei {\bf 345}, 21 (1993)}.

\bibitem{Grahn2009}
T.~Grahn et~al.,
\newblock \href {http://dx.doi.org/10.1103/PhysRevC.80.014324} {Physical Review
  C {\bf 80}, {014324} (2009)}.

\bibitem{Scheck2010}
M.~Scheck et~al.,
\newblock \href {http://dx.doi.org/10.1103/PhysRevC.81.014310} {Physical Review
  C {\bf 81}, {014310} (2010)}.

\bibitem{Deng1995}
J.~Deng et~al.,
\newblock \href {http://dx.doi.org/10.1103/PhysRevC.52.595} {Physical Review C
  {\bf 52}, 595 (1995)}.

\bibitem{Ma1993}
W.~Ma et~al.,
\newblock \href {http://dx.doi.org/10.1103/PhysRevC.47.R5} {Physical Review C
  {\bf 47}, R5 (1993)}.

\bibitem{Lee1990}
I.-Y. Lee,
\newblock \href {http://dx.doi.org/10.1016/0375-9474(90)91181-P} {Nuclear
  Physics A {\bf 520}, c641 (1990)}.

\bibitem{GSSORT}
T.~Lauritsen,
\newblock \href {http://www.phy.anl.gov/gammasphere/doc/GSSort/} {{GSSort
  (unpublished)}}.

\bibitem{Brun1997}
R.~Brun and F.~Rademakers,
\newblock \href {http://dx.doi.org/10.1016/S0168-9002(97)00048-X} {Nuclear
  Instruments and Methods in Physics Research Section A {\bf 389}, 81 (1997)}.

\bibitem{Petkov1994}
P.~Petkov,
\newblock \href {http://dx.doi.org/10.1016/0168-9002(94)90636-X} {Nuclear
  Instruments and Methods in Physics Research Section A: Accelerators,
  Spectrometers, Detectors and Associated Equipment {\bf 349}, 289 (1994)}.

\bibitem{Bohm1993}
G.~B\"{o}hm, A.~Dewald, P.~Petkov, and P.~von Brentano,
\newblock \href
  {http://www.sciencedirect.com/science/article/pii/016890029390944D} {Nuclear
  Instruments and Methods in Physics Research Section A: Accelerators,
  Spectrometers, Detectors and Associated Equipment {\bf 329}, 248 (1993)}.

\bibitem{Dewald2012}
A.~Dewald, O.~M\"{o}ller, and P.~Petkov,
\newblock \href {http://dx.doi.org/10.1016/j.ppnp.2012.03.003} {Progress in
  Particle and Nuclear Physics {\bf 67}, 786 (2012)}.

\bibitem{Napatau}
B.~Saha,
\newblock \href {http://www.ikp.uni-koeln.de/misc/doc/napa/} {{Napatau or
  Tk-Lifetime-Analysis (unpublished)}}.

\bibitem{Dewald1989}
A.~Dewald, S.~Harissopulos, and P.~Brentano,
\newblock \href {http://dx.doi.org/10.1007/BF01294217} {Zeitschrift f\"{u}r
  Physik A Atomic Nucleir Physik {\bf 334}, 163 (1989)}.

\bibitem{Petkov2001}
P.~Petkov, A.~Dewald, and P.~von Brentano,
\newblock \href {http://dx.doi.org/10.1016/S0168-9002(00)00797-X} {Nuclear
  Instruments and Methods in Physics Research Section A {\bf 457}, 527 (2001)}.

\bibitem{Proetel1974}
D.~Proetel, R.~M. Diamond, and F.~S. Stephens,
\newblock \href {http://dx.doi.org/10.1016/0370-2693(74)90653-4} {Physics
  Letters B {\bf 48}, 102 (1974)}.

\bibitem{Rud1973}
N.~Rud, D.~Ward, H.~Andrews, R.~Graham, and J.~Geiger,
\newblock \href {http://dx.doi.org/10.1103/PhysRevLett.31.1421} {Physical
  Review Letters {\bf 31}, 1421 (1973)}.

\bibitem{Ma1986}
W.~C. Ma et~al.,
\newblock \href {http://dx.doi.org/10.1016/0370-2693(86)90345-X} {Physics
  Letters B {\bf 167}, 277 (1986)}.

\bibitem{Kondev1999}
F.~Kondev et~al.,
\newblock \href {http://dx.doi.org/10.1103/PhysRevC.61.011303} {Physical Review
  C {\bf 61}, {011303} (1999)}.

\bibitem{Kondev2000}
F.~Kondev et~al.,
\newblock \href {http://dx.doi.org/10.1103/PhysRevC.62.044305} {Physical Review
  C {\bf 62}, {044305} (2000)}.

\bibitem{Mariscotti1969}
M.~A.~J. Mariscotti, G.~Scharff-Goldhaber, and B.~Buck,
\newblock \href {http://dx.doi.org/10.1103/PhysRev.178.1864} {Physical Review
  {\bf 178}, 1864 (1969)}.

\bibitem{Lane1995}
G.~J. Lane et~al.,
\newblock \href {http://dx.doi.org/10.1016/0375-9474(95)00080-K} {Nuclear
  Physics A {\bf 589}, 129 (1995)}.

\bibitem{Dracoulis1994}
G.~D. Dracoulis,
\newblock \href {http://dx.doi.org/10.1103/PhysRevC.49.3324} {Physical Review C
  {\bf 49}, 3324 (1994)}.

\bibitem{Dickmann1974}
F.~Dickmann and K.~Dietrich,
\newblock \href {http://dx.doi.org/10.1007/BF02126197} {Zeitschrift f\"{u}r
  Physik {\bf 271}, 417 (1974)}.

\bibitem{Walpe2012}
J.~C. Walpe et~al.,
\newblock \href {http://dx.doi.org/10.1103/PhysRevC.85.057302} {Physical Review
  C {\bf 85}, {057302} (2012)}.

\bibitem{Dracoulis1988}
G.~D. Dracoulis et~al.,
\newblock \href {http://dx.doi.org/10.1016/0375-9474(88)90244-8} {Nuclear
  Physics A {\bf 486}, 414 (1988)}.

\bibitem{Guttormsen1981}
M.~Guttormsen,
\newblock \href {http://dx.doi.org/10.1016/0370-2693(81)90998-9} {Physics
  Letters B {\bf 105}, 99 (1981)}.

\bibitem{Helppi1983}
H.~Helppi, S.~K. Saha, P.~J. Daly, S.~R. Faber, and T.~L. Khoo,
\newblock \href {http://dx.doi.org/10.1103/PhysRevC.28.1382} {Physical Review C
  {\bf 28}, 1382 (1983)}.

\bibitem{Delaroche1994}
J.~Delaroche et~al.,
\newblock \href {http://dx.doi.org/10.1103/PhysRevC.50.2332} {Physical Review C
  {\bf 50}, 2332 (1994)}.

\bibitem{Petts2009}
A.~Petts et~al.,
\newblock \href {http://dx.doi.org/10.1063/1.3087057} {{Lifetime Measurements
  and Coulomb Excitation of Light Hg Nuclei}},
\newblock in {\em AIP Conference Proceedings}, volume 1090, pages 414--418,
  AIP, 2009.

\bibitem{Bree2013}
N.~Bree et~al.,
\newblock Physical Review Letters (submitted)  (2014).

\end{thebibliography}

\end{document}